# Experimental discovery of bulk-disclination correspondence

Yang Liu[1,#], Shuwai Leung[2,#], Fei-Fei Li[2,#], Zhi-Kang Lin[1,#], Xiufeng Tao[2], Yin Poo[2,†], Jian-Hua Jiang[1,†]

[1]*School of Physical Science and Technology, and Collaborative Innovation Center of Suzhou Nano Science and Technology, Soochow University, 1 Shizi Street, Suzhou, 215006, China*

[2]*School of Electronic Science and Engineering, Nanjing University, Nanjing, 210093, China*

[#]These authors contributed equally to this work.

[†]Correspondence and requests for materials should be addressed to jianhuajiang@suda.edu.cn (Jian-Hua Jiang), ypoo@nju.edu.cn (Yin Poo).

## Abstract

**Most natural and artificial materials have crystalline structures from which abundant topological phases emerge [1-6]. The bulk-edge correspondence, widely-adopted in experiments to determine the band topology from edge properties, however, becomes inadequate in discerning various topological crystalline phases [7-17], leading to great challenges in the experimental classification of the large family of topological crystalline materials [4-6]. Theories predict that disclinations, ubiquitous crystallographic defects, provide an effective probe of crystalline topology beyond edges [18-21], which, however, has not yet been confirmed in experiments. Here, we report the experimental discovery of the bulk-disclination correspondence which is manifested as the fractional spectral charge and robust bound states at the disclinations. The fractional disclination charge originates from the symmetry-protected bulk charge patterns---a fundamental property of many topological crystalline insulators (TCIs). Meanwhile, the robust bound states at disclinations emerge as a secondary, but directly observable property of TCIs. Using reconfigurable photonic crystals as photonic TCIs with higher-order topology, we observe those hallmark features via pump-probe and near-field detection measurements. Both the fractional charge and the localized states are demonstrated to emerge at the disclination in the TCI phase but vanish in the trivial phase. The experimental discovery**

**of bulk-disclination correspondence unveils a novel fundamental phenomenon and a new paradigm for exploring topological materials.**

Symmetry plays a pivotal role in the study of topological states of matter [1-3]. Recent attempts for the complete catalogue of topological materials reveal that the 230 space groups in solid crystals lead to a vast number of TCIs with a very large classification [4-6]. The great challenge arising is how to identify the crystalline topology experimentally, bearing in mind that the bulk-edge correspondence becomes inadequate for such a purpose [11-17, 22-29]. This is primarily because crystalline surfaces have lower symmetry and thus less information than the bulk. For instance, in the recently discovered higher-order topological insulators with fractional dipole or quadrupole polarizations [11, 22-24, 27-29], the edges do not carry the full information of the bulk charge patterns.

Disclinations, topological defects commonly exist in crystals, have been predicted as an efficient *bulk* probe of various types of crystalline topology [18-21]. A primary topological observable is the fractional charge bound to disclinations which emerge due to the rotational disruption of the crystalline structure. Such a fractional charge can be used to deduce the symmetry-protected bulk charge patterns and thus serve as a distinctive observable for various TCIs. Note that although fractional charges at edges [16, 17] and dislocations [30-35] can probe some TCIs, but they can only probe the bulk charge polarization instead of the bulk charge patterns.

The bulk-disclination correspondence can be illustrated by a concrete example of two-dimensional (2D) TCI with the six-fold rotation, $C_6$, symmetry. As shown in Fig. 1a, the disclination is constructed by removing a $2\pi/6$ sector of the perfect crystal and gluing the remaining together. We consider a TCI that can be theoretically described by a tight-binding model illustrated in Fig. 1b with the inter-unit-cell coupling stronger than the intra-unit-cell coupling. The normal insulator (NI) phase is realized by interchanging these couplings. Both the TCI and the NI have well-defined Wannier centers that indicate the charge distribution in the bulk (Fig. 1b). In the TCI, the Wannier centers are at the edges of the unit-cell. Thus, each Wannier center is shared by two unit-cells and contributes $\frac{1}{2}$ charge to one of them. In the NI phase, the Wannier centers are at the center of the unit-cell.

The unique charge pattern of the TCI leads to a fractional charge, $\frac{1}{2}$ modulo 1, and the disclination states (see Figs. 1c and 1d) [18-21]. As illustrated in the inset of Fig. 1d, the boundary of the disclination core runs through five Wannier centers (dashed circles), which induces a separation of the charge associated with these Wannier centers and hence the formation of localized disclination states. The remaining Wannier centers (purple dots) contribute a fractional charge 2.5 to each unit-cell in the disclination (orange) region (Fig. 1e). Each unit-cell in the bulk (gray) region has an integer charge 3, whereas the charge per unit-cell in the edge (green) and corner (blue) regions deviates substantially from the bulk value. In contrast, such properties disappear in the NI where all unit-cells have the integer charge 3 (Fig. 1f). The fractional charge in the disclination region thus manifests the bulk-disclination correspondence (see Supplementary Note 1 for details). We remark that beside the examples studied in this work, the bulk-disclination correspondence is more general and applies to TCIs that are not Wannier representable [18-21], such as fragile topological insulators.

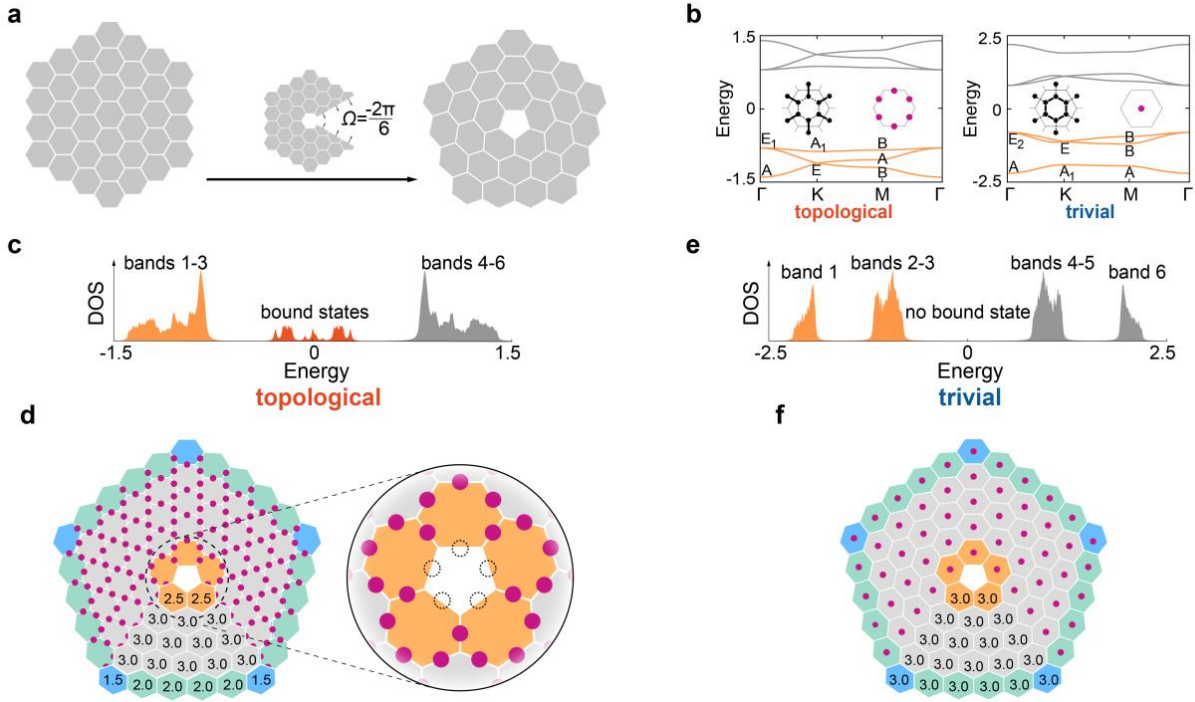

**Figure 1 | Disclination as a bulk probe of crystalline topology. a**, Illustration of the formation of a disclination in hexagonal crystals. **b,** Band structures and Wannier centers (red dots in right insets) of the topological and trivial tight-binding models (illustrated in the left insets; solid/dashed lines represent strong/weak couplings, respectively). Little group representations are labeled for the occupied bands 1-3.

**c-f:** DOS and spectral charge distribution in finite disclinations for the topological (**c-d**) and trivial (**e-f**) phases. The in-gap DOS in **c** originates from the bound states at the disclination, corners and edges. Numbers in **d** and **f** denote the spectral charges of the unit-cells in the disclination (orange), bulk (gray), edge (green) and corner (blue) regions for nearly 1/5 of the whole structure. The zoom-in inset of **d** illustrates the origin of the fractional charge and the disclination states. Results are calculated with strong (weak) coupling being -1 (-0.2).

We use 2D photonic crystals (PhCs) [36], artificial materials that can realize various photonic topological phases [37-44], to observe the bulk-disclination correspondence. The TCI and NI in Fig. 1 can be realized in 2D PhCs as illustrated in Fig. 2a (see Materials and Methods for details). The PhC can be configured by tuning the distance $d$, to switch between the TCI and NI phases (Figs. 2b-2d). In experiments, the TCI (NI) is realized with $d/a = 0.5$ (0.23). Their photonic bands are presented in Figs. 2b and 2c, respectively. We find that the Wannier centers of the TCI (NI) are at the edges (center) of the unit-cell (see Supplementary Note 2). The transition between the TCI and the NI takes place at $d = a/3$ [40, 43]. The photonic TCI is, in fact, a higher-order topological insulator [28-35] with gapped edge states, in-gap corner states and fractional corner charge (see Supplementary Note 3). However, these properties cannot completely determine the bulk topology, as there are similar higher-order topological states that have distinct bulk topology. In Supplementary Note 1, we show that all kinds of spinless TCIs with $C_6$ symmetry can be uniquely distinguished and identified through the bulk-disclination correspondence.

The crystalline topology and the bulk charge pattern can be deduced from the symmetry indicators [16, 21] (or equivalently the band representations [3]). The topological index is given by [16, 21] $\chi = (\chi_M, \chi_K)$ where the symmetry indicators are $\chi_M = \#M_1^{(2)} - \#\Gamma_1^{(2)}$ and $\chi_K = \#K_1^{(3)} - \#\Gamma_1^{(3)}$. Here, $\#\Pi_p^{(n)}$ is the number of bands below the band gap at a high-symmetry point $\Pi = \Gamma, M, K$ with the $C_n$ rotation eigenvalue $e^{i2\pi(p-1)/n}$ ($p = 1, ..., n$). We find that $\chi_M = -2$ or 0 for the TCI and NI, respectively, whereas $\chi_K = 0$ for both (see Supplementary Note 3). Remarkably, the fractional disclination charge, $Q_{dis}$, is determined by the topological index of the TCI as [21]

$$Q_{dis} = \frac{\Omega}{2\pi}\left(\frac{3}{2}\chi_M - \chi_K\right) \bmod 1, \tag{1}$$

where $\Omega$ is the Frank angle (Fig. 1a). The above equation, which elegantly manifests the bulk-disclination correspondence, yields $Q_{dis} = \frac{1}{2}$ for the TCI and $Q_{dis} = 0$ for the NI, respectively.

Although photons are neutral particles, the "fractional charge" can be revealed from the local density-of-states (LDOS) measurements. By integrating the measured LDOS up to the bulk band gap, the number of modes per unit-cell, denoted as the "spectral charge", can be extracted (see Materials and Methods for details). In the electronic counterparts, electron filling of eigenstates up to the band gap gives exactly the same charge in unit of the elementary charge $e$. To test this prediction, we first numerically calculate the spectral charge distributions for the photonic TCI and NI disclinations (see Supplementary Note 4 for details of the calculation). The results in Figs. 2e-2f show good agreement with the theoretical predictions in Figs. 1d and 1f.

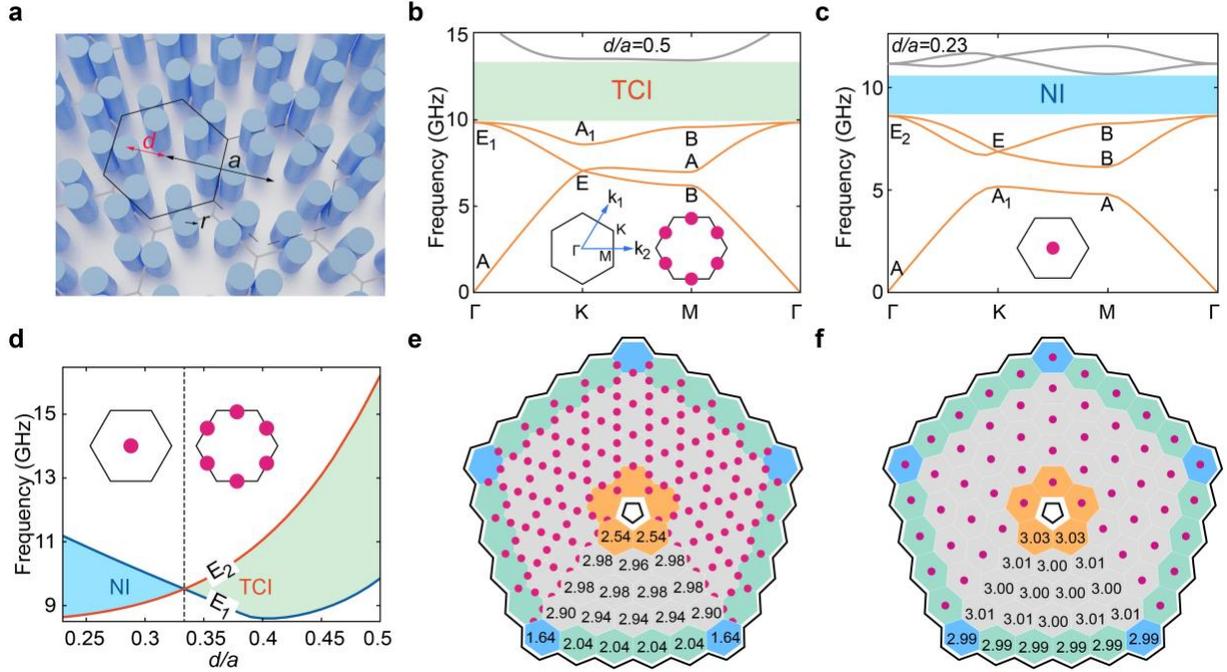

**Figure 2 | Photonic TCI and disclination. a,** Illustration of a 2D topological PhC. In each unit-cell, there are six dielectric pillars with identical radius $r = 2$ mm. Lattice constant is $a = \sqrt{3}$ cm. Distance between the unit-cell center and the center of a dielectric pillar is $d$. **b-c,** Photonic bands of PhCs with $d/a = 0.5$ (**b**) and 0.23 (**c**). Little group representations are labeled at the high-symmetry points. Insets illustrate the

Brillouin zone and the Wannier centers. **d**, Topological 'phase diagram' of the PhC. Transition between the TCI and NI is indicated by the band-crossing between the $E_1$ and $E_2$ representations at the Γ point. **e-f**, Spectral charge (numbers) and Wannier center (purple dots) distributions for the topological and trivial PhCs, respectively. Each disclination structure include 75 unit-cells. The black lines outside and inside denote the perfectly reflecting boundaries (see Materials and Methods for details).

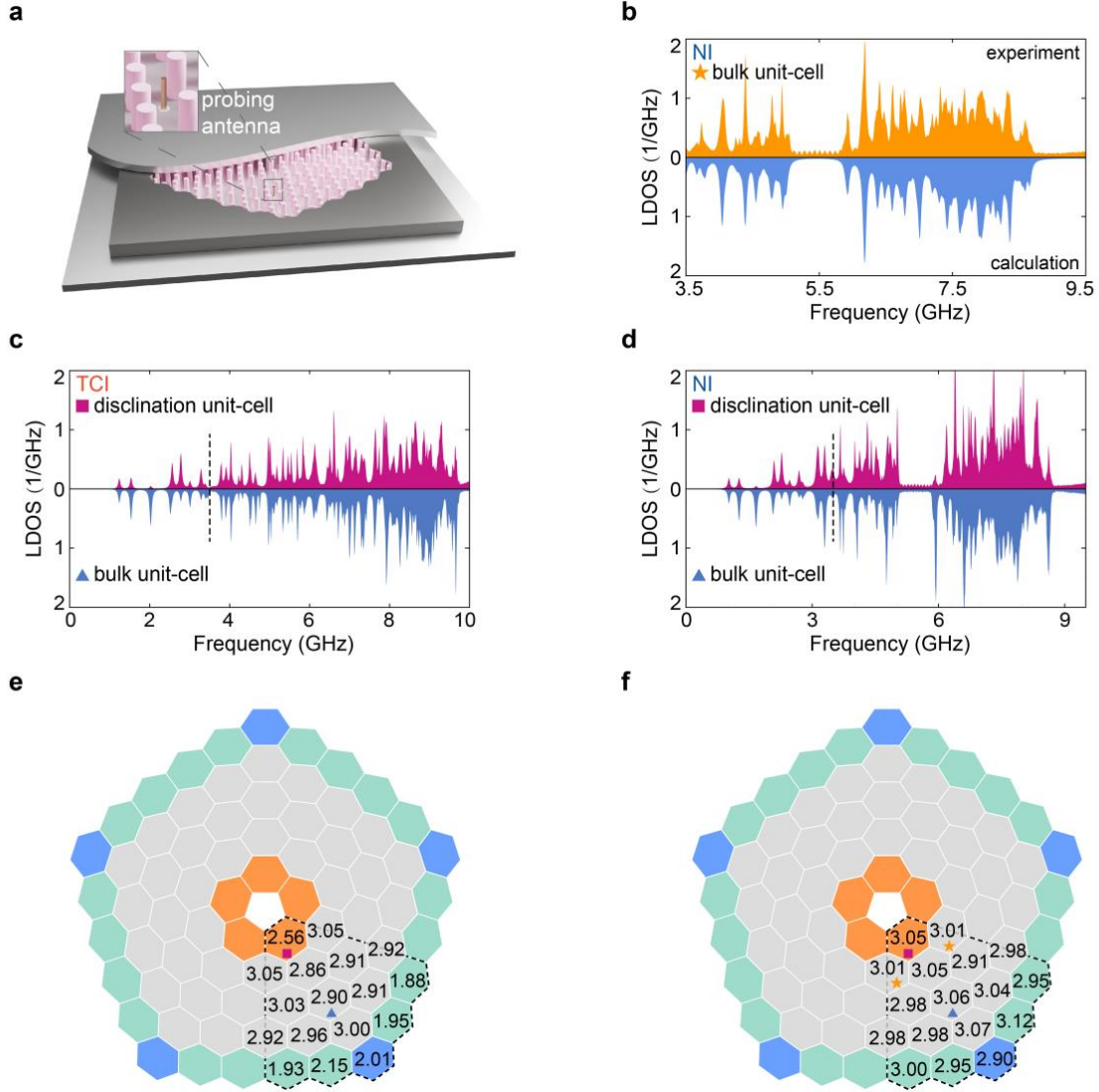

**Figure 3 | Bulk-disclination correspondence. a**, Experimental set-up for the measurement of the LDOS through the reflection spectroscopy of a short probing antenna. **b**, Measured (orange) and calculated (blue) LDOS of two half unit-cells together (orange stars in **f**) in the NI disclination. **c-d**, Measured LDOS of the disclination (purple square) and bulk (dark-blue triangle) unit-cells for the TCI and NI disclinations, respectively. **e-f**, Measured spectral charges of the unit-cells in 1/5 of the disclination structure (enclosed by the dashed lines) for the TCI and NI PhCs, respectively. The LDOS's are presented up to a frequency
6

in the band gap (10 GHz for the TCI and 9.5 GHz for the NI). The LDOS's below 3.5 GHz (vertical dashed lines in **c-d**), which is unattainable from our instruments, comes from calculations.

To obtain the spectral charge, we measure the LDOS through the classical-version of the Purcell effect which states that the radiation resistance of a subwavelength antenna is proportional to the photonic LDOS [44]. The antenna's radiation resistance is determined from the reflection spectroscopy of a coaxial cable port connecting to the antenna with a frequency resolution of 1 MHz (Fig. 3a; see Materials and Methods for details). By comparing the radiation resistance in our systems and that in a reference photonic system with known DOS (see Supplementary Note 5), we extract the LDOS in the disclinations. Due to the five-fold rotation symmetry of the disclination, the measurement is performed only for 1/5 of the disclination structure.

As shown in Fig. 3b, the measured LDOS agrees well with the calculated LDOS. In Figs. 3c and 3d, we present the measured LDOS of a disclination unit-cell together with that of a bulk unit-cell in Figs. 3c and 3d for the TCI and NI cases, respectively. The difference between the disclination and bulk LDOS is larger in the TCI than for the NI, since the former yields a spectral charge of 2.5 for the disclination unit-cell (lower than the bulk value of 3). The spectral charges deduced from the LDOS measurements are summarized in Figs. 3e and 3f for the TCI and NI, respectively. For the TCI, the spectral charge per unit-cell is close to 2.5 (3) for the disclination (bulk) region, while it is close to 2 in the edge region. These results agree with the theory and simulation in Figs. 1 and 2. The deviation of the corner charge from the theoretical value is possibly due to uncontrollable sample deformations in the experiment. In comparison, for the NI, the spectral charge is close to the bulk value 3 in every unit-cell.

In addition to the fractional disclination charge, we also observe experimentally the localized disclination states as a secondary effect of the bulk-disclination correspondence. The disclination states are measured through the pump-probe transmission and near-field scanning methods which involve a setup with multiple antennas as illustrated in Fig. 4a. We first reveal the existence of disclination states through the pump-probe transmission. The source and detector antennas (labeled as A and B in Fig. 4a, respectively) are placed at the opposite sides of the disclination to probe the disclination states. For the TCI disclination, three resonances (denoted as $\alpha$, $\beta$ and $\gamma$, separately) are observed in the band gap region which agrees well with the calculated photonic spectrum and the simulated transmission (Fig. 4b). Calculation reveals that the $\alpha$ resonance is nondegenerate, while both the $\beta$ and $\gamma$ resonances are doubly degenerate.

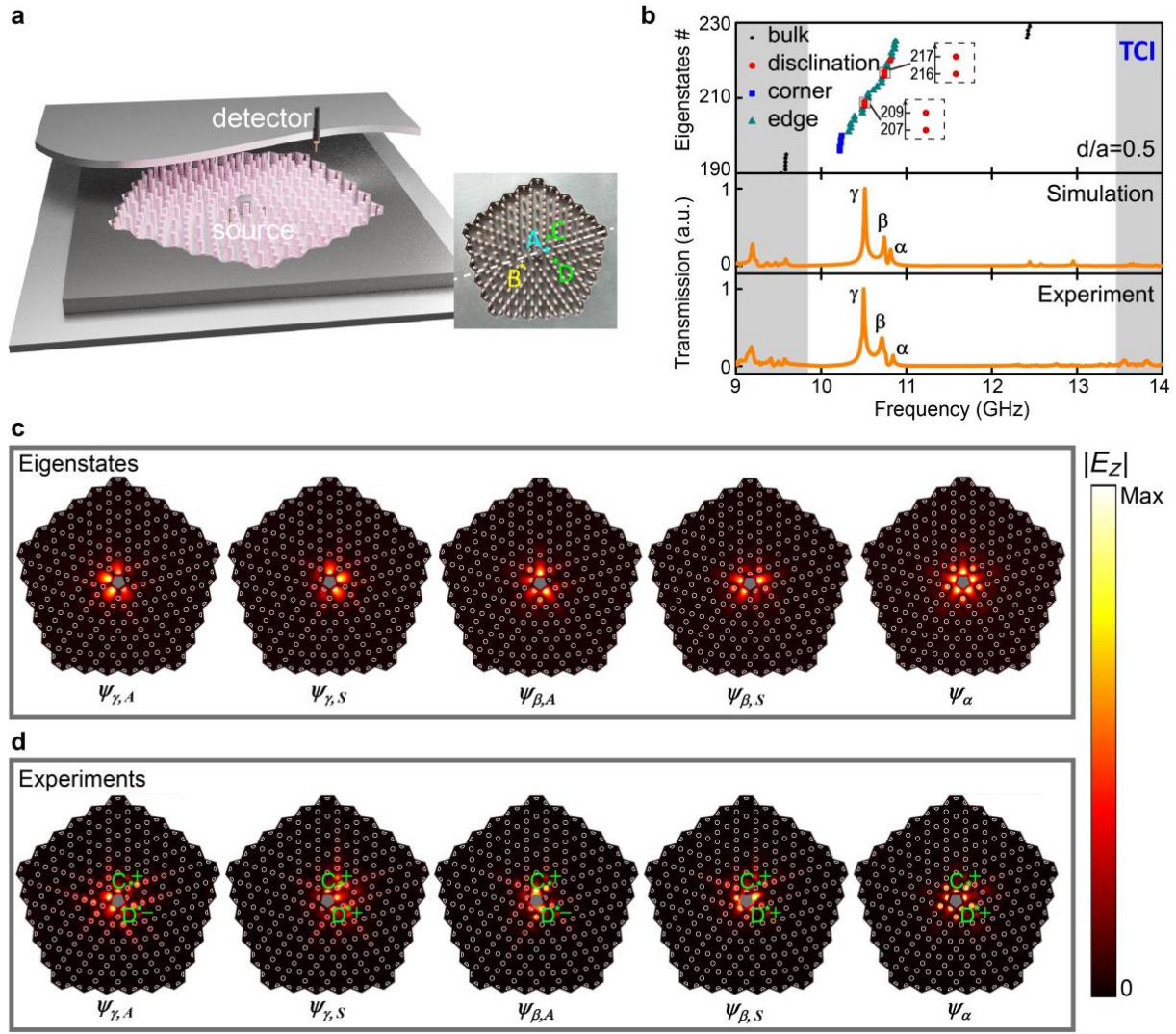

**Figure 4 | Measurement of the disclination states. a**, Illustration of the experimental setup. A PhC ($d = a/2$) disclination made of Al$_2$O$_3$ pillars (pink) is cladded by parallel metal plates (gray) above and below, and is surrounded by aluminum frame (dark gray). Lower-right: top view of the structure. **b**, Eigenstates spectrum (top), and the simulated (middle) and measured (down) transmission between source A and detector B. Gray regions denote the bulk-band regions. **c-d**, Electric-field profiles of disclination states

from eigenstates calculation (**c**) and measurements (**d**). The + and − signs indicate whether the sources C and D (green stars) have the same sign or opposite signs.

The wavefunctions of the disclination states are measured using the resonant near-field scanning (see Materials and Methods). The challenge is to measure the doubly degenerate states which can hybridize with each other in arbitrary manners. A single point-like source can excite only one hybridized state (the 'bright-state') and leave the other one (the 'dark-state') undetectable. We design two point-like sources, C and D, which locate symmetrically around the mirror symmetry plane of the disclination (labeled by the white-dashed line in Fig. 4a) to overcome this challenge. With respect to the mirror symmetry, the doubly degenerate states form the symmetric and anti-symmetric states which can be excited, respectively, when C and D have the phase difference 0 or $\pi$. By tuning the phase difference and the frequency of the sources, we can selectively excite all the five disclination states, $\psi_\alpha$, $\psi_{\beta,S}$, $\psi_{\beta,A}$, $\psi_{\gamma,S}$ and $\psi_{\gamma,A}$, and measure their wavefunctions using the near-field scanning method. The measured wavefunctions agree well with the calculated eigenstates wavefunctions (Figs. 4c and 4d show the amplitude, while the phase of the wavefunction is shown in Supplementary Note 6). These disclination states survive even when the boundary conditions are strongly modified (see Supplementary Note 7).

In contrast, in the NI disclination, we find from both calculation and experiments that there is no localized state in the band gap (see Supplementary Note 8 for details). This observation indicates that under certain boundary conditions, disclination states emerge as a side-effect of the bulk-disclination correspondence. Moreover, as demonstrated numerically in Supplementary Note 9, the photonic disclination states can serve as robust, subwavelength cavities which are valuable for future photonic applications such as lasing [45] and quantum emitters [43].

We discover experimentally a novel topological phenomenon, the bulk-disclination correspondence, through the detection of the fractional spectral charge and localized states at disclinations. Experimental identification of crystalline topology through these properties can be directly generalized to electronic topological crystalline materials where the disclination states and the LDOS can be measured directly using scanning tunneling microscopes. Furthermore, such experiments are facilitated by the common existence of disclinations in solid state materials. Our study thus opens a new regime for the study of topological crystalline phases---a frontier with a vast number of unexplored materials and phenomena.

## Materials and Methods

## Samples

The experimental system is based on quasi-2D photonic systems of transverse-magnetic polarization (electric field perpendicular to the 2D plane) created by metal plates cladding from above and below. The distance between the parallel metal plates is about 11 mm (its specific value varies slightly for different measurements). Samples are constructed by carefully pasting dielectric cylinders into a pre-printed drawing on which the location of each cylinder is marked. The dielectric cylinders and half-cylinders are made of commercial alumina ceramics ($Al_2O_3$) doped with chromium dioxide. These cylinders and half-cylinders have the same height of 10 mm and the same radius of 2 mm. The relative permittivity of the alumina ceramics is measured by the transmission/reflection method using the Weir-Nicolson-Ross method. It changes little at the X-band and can be treated as a constant of $8.4 - 0.02i$ for the frequency range in this work.

Each disclination structure includes 75 unit-cells as depicted in Figure 2. To prevent photons from radiating into the free space, we set perfectly reflecting boundaries (PRBs) at the outside of the disclinations. The PRBs are realized by aluminum frames in experiments. We also set the PRBs in the disclination core to regularize the spectral charges of the unit-cells close to the core. These PRBs suppress the non-Hermitian effect and make the spectral charge well-defined. As a consequence, the spectral charges of the unit-cells in the NI disclination are close to the theoretical value of 3 (consistent with the fact that there are three bands below the band gap).

## Experimental setup

The photonic LDOS, directly related to the antenna's Purcell factor, is determined by the one-port reflection spectroscopy of the coaxial cable port connecting to the antenna. A vector network analyzer Agilent PNA-X N5247A is used to excite the electromagnetic wave and record the magnitude and phase of the reflected electromagnetic wave received by the cable. In order to measure the antenna's Purcell factor, the length of the antenna must be much smaller than the photonic wavelength in free space [44]. In experiments, the length of the antenna is optimized to fulfill this requirement and to maintain considerable radiation efficiency (Antennas that are too short have very low radiation efficiency). We divide the measured frequency range 3.5-10 GHz into two regions: 3.5-5 GHz and 5-10 GHz where antennas of length 4.9 mm and 3.3 mm are adopted, respectively. A reference photonic system with uniform 2D photonic DOS is used to determine the LDOS in disclinations. The reflection spectroscopy is measured with a frequency resolution of 1 MHz. To determine the spectral charge of a unit-cell, the LDOS is measured in 24 small divisions of the unit-cell. The spectral charge is then obtained by integrating the LDOS over a unit-cell for the frequency range from 0 to a frequency in the band gap (i.e., 10 GHz in TCI and 9.5 GHz in NI). More details are presented in Supplementary Note 5.

For the measurement of the disclination states, a 1 mm air gap between the PhC and the upper metal plate is introduced to enable the moving of the lower metal plate where the PhC is mounted on. In the scanning measurements of the wavefunctions, the disclination states are excited resonantly by two source antennas (diameter 0.53 mm, labeled as C and D in Fig. 4a) with phase difference 0 or $\pi$. The two antennas have the same height of 9.9 mm and are inserted through two holes on the lower plate. In the scanning measurements, the lower plate (and the source antennas on it) is mounted on a translational stage with the scan step of 2 mm, while the upper plate and the detector antenna (which is embedded in the upper metal plate and connected to the detector cable) is kept stationary. The transmission spectra are obtained in a stationary setup without the air gap. In the transmission measurements, the source and the detector (both have a diameter 1.24 mm) are fixed. The frequency dependence of the transmission is quantified through the intensity of the detection signal, while the intensity of the input signal is fixed.

## Simulation

Numerical simulations are performed using a commercial finite-element simulation software (COMSOL MULTIPHYSICS) via the RF module. Simulations are performed for 2D transverse-magnetic harmonic modes (i.e., the electric-field is perpendicular to the 2D plane, along the *z* direction). The bulk band regions (gray) in all figures are calculated for infinite

PhCs, while the disclination states are calculated using the disclination structures with 30 unit-cells depicted in Figs. 3 and 4.

## Acknowledgements

Y.P, F.F.L, S.L, and X.T thank the National Natural Science Foundation of China (NSFC) (61671232 and 61771237), the Project Supported by the Fundamental Research Funds for the Central Universities (14380160 and 14380147) and the Project Funded by the Priority Academic Program Development of Jiangsu Higher Education Institutions. Y.L, Z.K.L and J.H.J are supported by the Jiangsu Province Specially-Appointed Professor Funding, the National Natural Science Foundation of China (Grant No. 11675116) and the Project Funded by the Priority Academic Program Development of Jiangsu Higher Education Institutions.

## Author contributions

J.H.J initiated the project. J.H.J and Y.P guided the research. J.H.J and Y.L established the theory. Y.L performed the numerical calculations and simulations. Z.K.L and Y.P are involved in some of the theory and simulations, respectively. F.F.L, S.L, X.T, J.H.J and Y.P designed and achieved the experimental set-up and the measurements. All the authors contributed to the discussions of the results and the manuscript preparation. J.H.J, Y.P and Y.L wrote the manuscript and the Supplementary Information.

## Competing Interests

The authors declare that they have no competing financial interests.

## Data availability

All data are available in the manuscript and the Supplementary Information. Additional information is available from the corresponding authors through proper request.

## Code availability

We use the commercial software COMSOL MULTIPHYSICS to perform the electromagnetic simulations and eigenstates calculations. Request to computation details can be addressed to the corresponding authors.